# Heliyon

# Multi-technique analysis of precipitable water vapor estimates in the sub-Sahel West Africa

**Oluwasesan A. Falaiye [a], Oladiran J. Abimbola [b,\*], Rachel T. Pinker [c], Daniel Pérez-Ramírez [d,e], Alexander A. Willoughby [f]**

[a] *University of Ilorin, Ilorin, Kwara State, Nigeria*

[b] *Federal University Lafia, Lafia, Nasarawa State, Nigeria*

[c] *Department of Atmospheric and Oceanic Science, University of Maryland, College Park, MD, USA*

[d] *Mesoscale Atmospheric Processes Laboratory, NASA Goddard Space Flight Center, Greenbelt, MD, USA*

[e] *Goddard Earth Sciences Technology and Research, Universities Space Research Association (GESTAR/USRA), Columbia, MD, USA*

[f] *Department of Physical Sciences, Covenant University, Sango-Otta, Ogun State, Nigeria*

\* Corresponding author.

E-mail address: ladiran@gmail.com (O.J. Abimbola).

## Abstract

Precipitable water vapor (PWV) is an important climate parameter indicative of available moisture in the atmosphere; it is also an important greenhouse gas. Observations of precipitable water vapor in sub-Sahel West Africa are almost non-existent. Several Aerosol Robotic Network (AERONET) sites have been established across West Africa, and observations from four of them, namely, Ilorin (4.34° E, 8.32° N), Cinzana (5.93° W, 13.28° N), Banizoumbou (2.67° E, 13.54° N) and Dakar (16.96° W, 14.39° N) are being used in this study. Data spanning the period from 2004 to 2014 have been selected; they include conventional humidity parameters, remotely sensed aerosol and precipitable water information and numerical model outputs. Since in Africa, only conventional information on humidity parameters is available, it is important to utilize the unique observations from the AERONET network to calibrate empirical formulas frequently used to estimate precipitable water vapor from humidity measurements. An empirical formula of the form $PWV = aT_d + b$ where $T_d$ is the surface dew point temperature, $a$ and $b$ are constants, was fitted to the data and is proposed as applicable to the climatic condition of the sub-Sahel. Moreover, we have also used the AERONET information to evaluate the capabilities of well-established numerical weather prediction (NWP) models such as ERA Interim Reanalysis, NCEP-DOE Reanalysis II and NCEP-CFSR, to estimate precipitable water vapor in the sub-Sahel West Africa; it was found that the models tend to overestimate the amount of precipitable water at the selected sites by about 25 %.

Keywords: Atmospheric science, Environmental science, Earth sciences

## 1. Introduction

Atmospheric precipitable water vapor (PWV) (the height of liquid water obtained if all the water vapor in an atmospheric column over a unit area is condensed) plays an important role in the hydrological cycle as it is formed by evaporation/evapotranspiration from the surface into the atmosphere, can condense into clouds and may return back to the surface in the form of precipitations. The latent heat of vaporization, which is released whenever atmospheric water vapor condenses, is an important aspect of the atmospheric energy budget providing diabatic heating and driving local and global weather systems (Trenberth et al., 2007).

The ability of water molecules to warm the atmosphere by absorbing and re-emitting radiation makes water vapor an important component of greenhouse gases and its effect on climate change processes is of interest. Water vapor's ability to absorb and re-emit electromagnetic waves has a profound effect on the propagation of radio waves in the atmosphere. A CIMEL CE-318-4 instrument is a sun-photometer used in the Aerosol Robotic Network (AERONET) (Holben et al., 1998) with a primary focus on estimating aerosol optical depth. Additionally, it works in the water-vapor absorption band around 940 nm that allows retrieving of PWV. Detailed information on this instrument, its calibration and utilization can be found in (Holben et al., 1998, 2001). The data used here are from level 2.0 of the CIMEL observations (Smirnov et al., 2000, 2004).

In areas such as Africa, information on the moisture content of the atmosphere is most frequently obtained from numerical weather prediction models or measurements of humidity using an extensive number of empirical formulas. The AERONET's CIMEL observations in West Africa provide an opportunity to assess the performance of empirical formulas under the environmental conditions of sub-Sahel as well as the performance of well-established numerical weather prediction models. Data used will be

 

described in Section 2, methodology will be described in Section 3, results will be presented in Section 4, and conclusions will be summarized in Section 5.

## 2. Study area

### 2.1. Ground observations

The AERONET's CIMEL sun-photometers across West Africa are located at the following sites: Ilorin, Nigeria (8.32° N, 4.34° E, 350 *m* amsl), Dakar, Senegal (14.39° N, 16.96° W, 0 *m* amsl), Banizoumbou, Niger (13.54° N, 2.67° E, 250 *m* amsl), Cinzana, Mali (13.28° N, 5.93° *W*, 285 *m* amsl), Ouagadougou, Burkina Faso (12.20° N, 1.40° *W*, 290 *m* amsl) and Djougou, Republic of Benin (9.76° N, 1.60° E, 400 *m* amsl) (Fig. 1). The methodology to derive precipitable water vapor from the CIMEL observations will be detailed in Section 3.1.

The surface meteorological data for each of the AERONET stations were obtained from the www.weatherspark.com weather services: surface meteorological data were obtained for a period of eleven years (2004−2014). The methodology to derive precipitable water vapor from humidity observations will be presented in Section 3.2.

### 2.2. Precipitable water vapor from numerical weather prediction models

Independent model results on precipitable water vapor that were selected for evaluation represent several well-known model prototypes. The ECMWF (http://www.ecmwf.int/) ERA Interim Reanalysis model (Berrisford et al., 2009) assimilates a variety of sources of meteorological parameters but clouds are produced internally. Similar to ERA Interim, the National Center for Environmental Predictions (NCEP) and the Department of Energy (DOE), NCEP-DOE Reanalysis II (Kistler et al., 2001) assimilate meteorological parameters from a variety of sources. The spatial resolution of the NCEP-DOE II data is at the T62 Gaussian Grid. Both ERA Interim and NCEP-DOE Reanalysis II use the Rapid Radiative Transfer Model (RRTM) developed by the Atmospheric and Environmental Research (AER) group (Mlawer et al., 1997). The National Centers for Environmental Prediction (NCEP) Climate Forecast System Reanalysis (CFSR) (Saha et al., 2010) is also used.

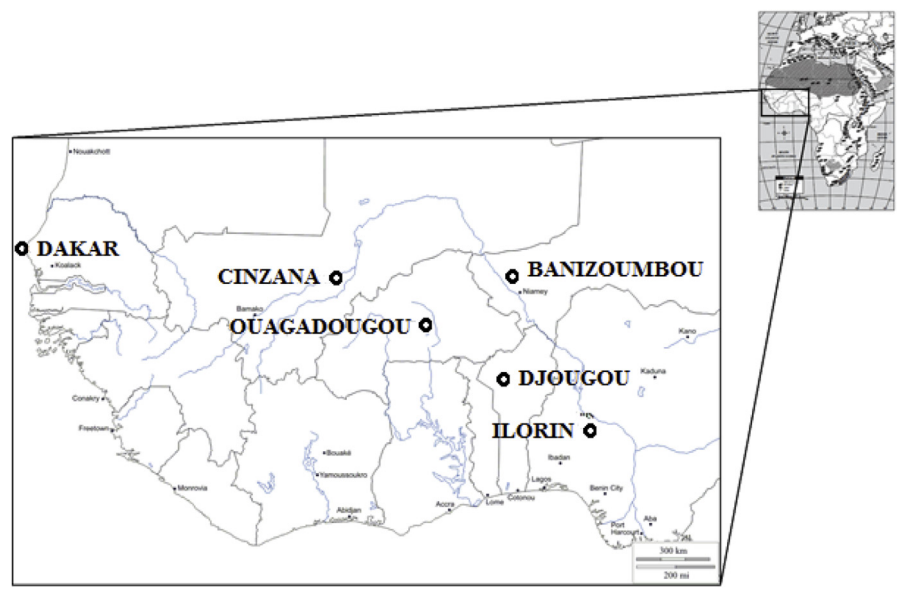

**Fig. 1.** Locations of the AERONET stations in West Africa.

## 3. Methodology

### 3.1. Derivation of precipitable water vapor from CIMEL observations

The sun-photometer method relies on the interaction of the solar electromagnetic energy with the atmospheric constituents before the energy reaches the earth surface. This interaction leads to scattering and absorption from which the amount of atmospheric water could be deduced. Particularly, in the near infrared spectrum, around 940 nm, there is a strong wavelength-dependent absorption by water vapor and the response of the instrument.

V (940 nm) to light in this spectral region is given by:

$$V(940\,nm) = V_o(940\,nm)d^{-2}\exp(-m_r\delta_{atm}(940\,nm))T_w(940\,nm) \tag{1}$$

where $V_O$ (940 nm) is the instrument calibration constant (signal that the instrument would measure if it were placed outside of the atmosphere), $d$ is the Earth-Sun distance (in astronomical units) at the time of observation, $m_r$ is the relative optical air mass, $\delta_{atm}$ (940 nm) is the total atmospheric optical depth (excluding absorption by water vapor) and $T_w$ (940 nm) is the water vapor transmittance around the 940 nm absorption bands. The computation of $V_O$ (940 nm) and $\delta_{atm}$ (940 nm) is done following AERONET procedures (Holben et al., 1998). For a straightforward retrieval of PWV, AERONET uses a simplified expression of $T_w$ (940 nm) given by (Reagan et al., 1987; Bruegge et al., 1992):

$$T_w(940\,nm) = \exp\left(-a(m_w PWV)^b\right) \tag{2}$$

where $m_w$ is the relative optical water vapor air mass and $a$ and $b$ are coefficients that depends on the wavelength position, width and shape of the sun-photometer filter function, and the atmospheric condition. Each AERONET instrument has its own unique set of 'a' and 'b' values depending on the filter configuration. These coefficients are considered fixed until the filter is changed. More information about the computation of coefficients 'a' and 'b' is in Smirnov et al. (2004).

 

## 3.2. Derivation of precipitable water from humidity observations

An approach to relate the natural logarithm of precipitable water vapor $lnPWV$ to dew point temperature $T_d$ (in °C) was suggested by Reitan (1963), Ojo (1970), Maduekwe and Ogunmola (1997), Okulov et al. (2002), Utah and Abimbola (2006) and Maghrabi and Al Dajani (2012):

$$lnPWV = k_1 T_d + k_2 \tag{3}$$

where $k_1$ and $k_2$ are constants.

More advanced methods for estimating precipitable water vapor include the use of radiosonde/dropsonde data (Willoughby et al., 2008; Adeyemi, 2008; Chang-Geun et al., 2012; Adeyemi and Joerg, 2012), microwave radiometers (Han et al., 1994), star photometers (Pérez-Ramírez et al., 2012), Raman lidars (Whiteman et al., 1992), Fourier transform spectrometers (Leblanc et al., 2011) and GPS/satellite data (Bevis et al., 1992, 1994; Ware et al., 1997; Jade et al., 2005; Sharifi et al., 2015; Li et al., 2018) which use the principle of tropospheric delays.

## 4. Results

## 4.1. Empirical models to estimate precipitable water vapor

A linear plot of the natural logarithm of precipitable water vapor calculated from the surface data $ln(PWV)$ (PWV in centimeter), for the year 2004−2014, is shown in Fig. 2. Within 95% confidence bound, the linear model was found to be given as

$$lnPWV = 0.055(\pm 0.001)T_d + 0.068(\pm 0.011) \tag{4}$$

The sum squared error (SSE) was found to be 882.47, while the root mean square error (RMSE) was found to be 0.31; SSE and RMSE results show good linear relation between the natural log of PWV and $T_d$, and this is further clarified by the value of the coefficient of determination ($R^2$) given by 0.727.

A plot of precipitable water vapor PWV, against the square of the surface relative humidity divided by the surface temperature [(RH)/T] is shown in Fig. 3. The linear model obtained results in:

$$PWV = 13.44\,[\pm 0.20]\left(\frac{RH}{T}\right) + 0.58[\pm 0.03] \tag{5}$$

Within the 95% confidence interval, SSE = 3185.97; RMSE = 0.71; $R^2$ = 0.706. The slope of Eq. (5) was found to range from 13.23 to 13.64; the intercept ranges between 0.54 and 0.61.

 

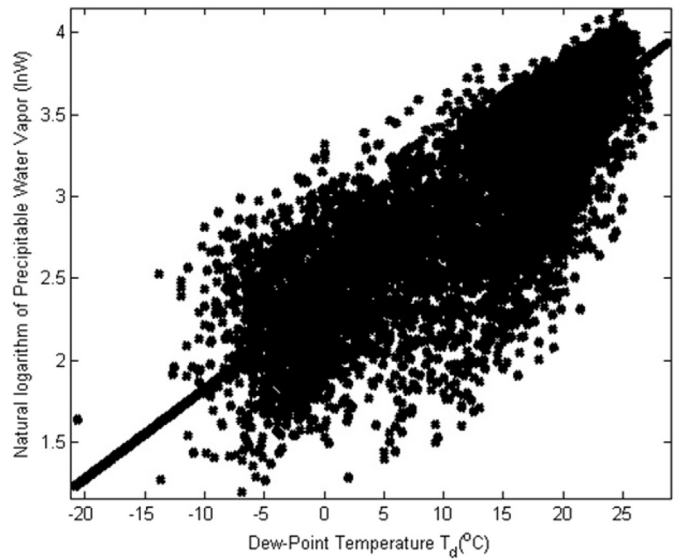

**Fig. 2.** Relationship between precipitable water vapor and surface dew-point temperature.

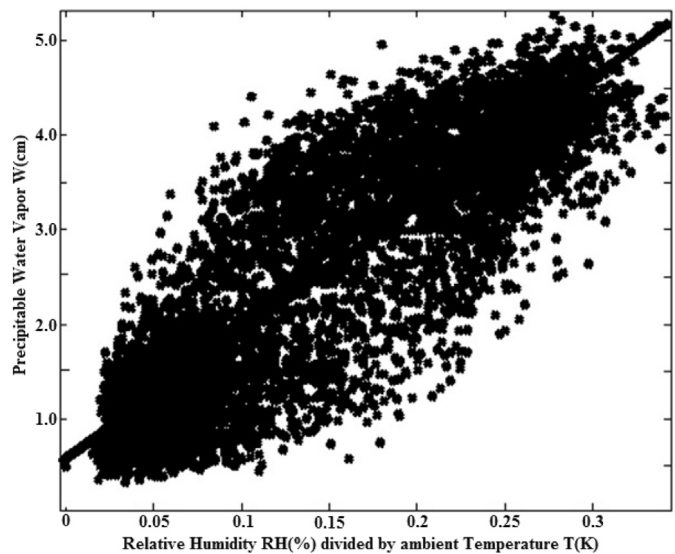

**Fig. 3.** Relationship between precipitable water vapor and relative humidity.

Fig. 4, shows a plot of precipitable water vapor against surface vapor pressure (*mbar*) divided by the surface ambient temperature (*K*) [e/T]. A linear fit was found most suitable for the distribution in Fig. 4; the linear fit resulted in:

$$PWV = 39.29[\pm 0.52]\left(\frac{e}{T}\right) + 0.25[\pm 0.03] \tag{6}$$

The goodness of fit statistics for Eq. (6) was found to be: SSE = 4742.00; RMSE = 0.72; $R^2$ = 0.694. The slope of Eq. (6) is in the range of 38.76−39.80, while the intercept is in the range of 0.21−0.28.

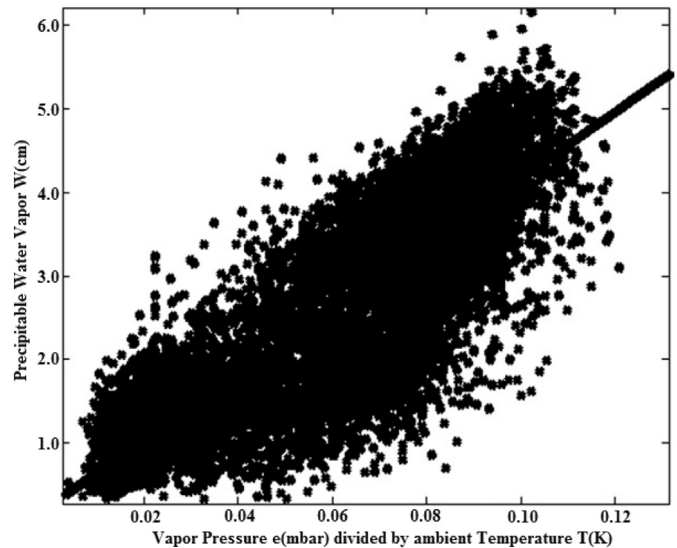

**Fig. 4.** Relationship between precipitable water vapor and a function of temperature (*T*) and vapor pressure (*e*) [i.e., *f*(*e*, *T*)].

## 4.2. Evaluation of model estimates of PWV against AERONET

PWV data from the AERONET's stations at Ouagadougou (12.20° N, 1.40° W) and Djougou (9.76° N, 1.60° E) as well as the NWP models of NCEP Reanalysis 2, NCEP-CFSR and ERA interim were used to evaluate the empirical models of Eqs. (4), (5), and (6). The validation results are summarized in Table 2 and the time series plot of the PWV derived from each empirical model together with the PWV from the AERONET's sun-photometer is shown in Fig. 5, for each of the stations in Ouagadougou and Djougou. From Table 2 as well as Fig. 5, it could be observed that the PWV ($T_d$), i.e., Eq. (4) generally has the best performance while PWV (e, T), i.e., Eq. (4) comes closer. As could be observed from Table 2, the comparison performance is better for the AERONET's data than for the NWP, and among the NWP models the worst performance is found between the NCEP Reanalysis 2 and the empirical models; this could be attributed to the grid resolution of each of the NWP model.

The precipitable water vapor data obtained from the CIMEL sun-photometer observations at Ilorin, Dakar, Banizoumbou and Cinzana were used to evaluate same parameter as derived from the three numerical weather prediction models described in Section 2.2 (ERA Interim, NCEP-DOE Reanalysis II, and CFSR). The quality of precipitable water vapor retrievals from the CIMEL instrument has been amply evaluated (Halthore et al., 1997; Pérez-Ramírez et al., 2014). For instance, Pérez-Ramírez et al. (2014) compared AERONET precipitable water vapor retrievals against radiosonde observations and other ground-based retrieval techniques such as microwave radiometry (MWR) and Global Positioning System (GPS)

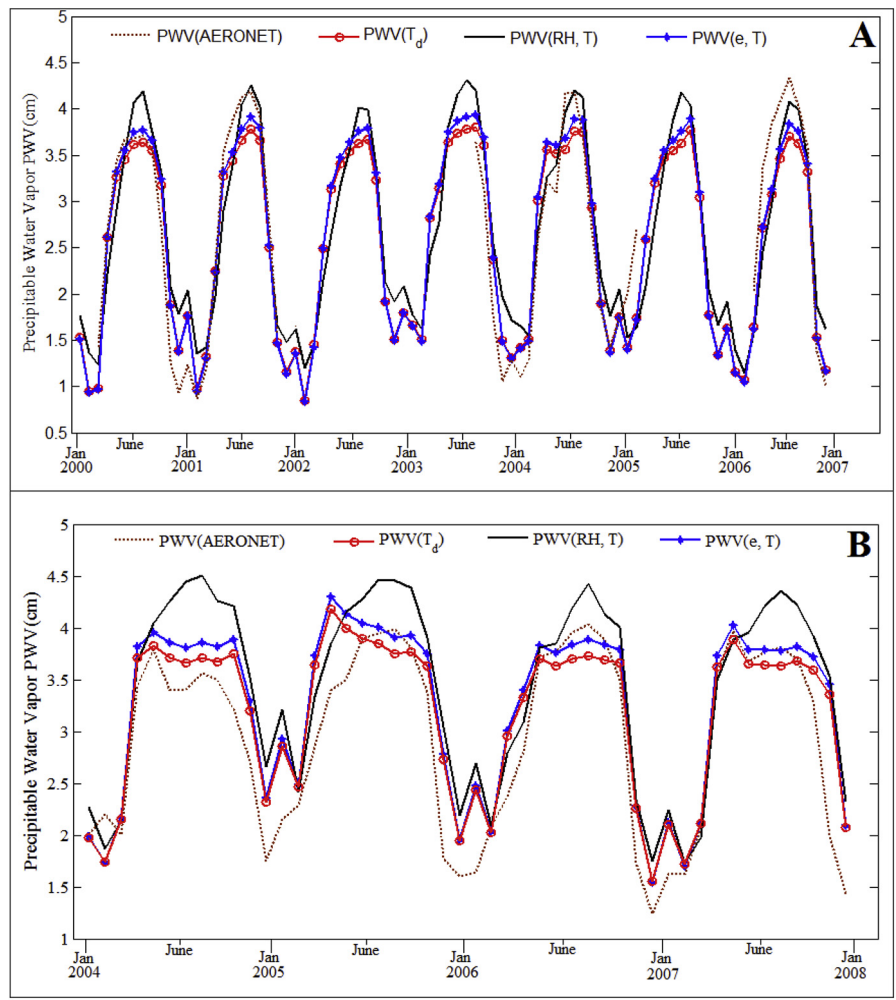

**Fig. 5.** Time Series of Precipitable Water Vapor from the empirical models of Eqs. (5), (6), and (7) with the AERONET's sun-photometer data for (A) Ouagadougou and (B) Djougou AERONET stations.

observations, it was found that the precipitable water vapor obtained by AERONET was lower than what was obtained by MWR and GPS by about 6.0−9.0% and about 6.0−8.0%, respectively. The AERONET values were also lower by approximately 5% than those obtained from numerous balloon-born radiosondes. These results point towards a consistent dry-bias in the retrievals of precipitable water vapor by AERONET although the differences are within the 10% systematic uncertainty estimated for the AERONET retrievals. As evident from Fig. 6 and as summarized in Table 1, results from ERA Interim and NCEP/CFSR are close to each other during the entire record, during the dry season and are also close to observations. During the summer period, all models overestimate precipitable water vapor more so NCEP/

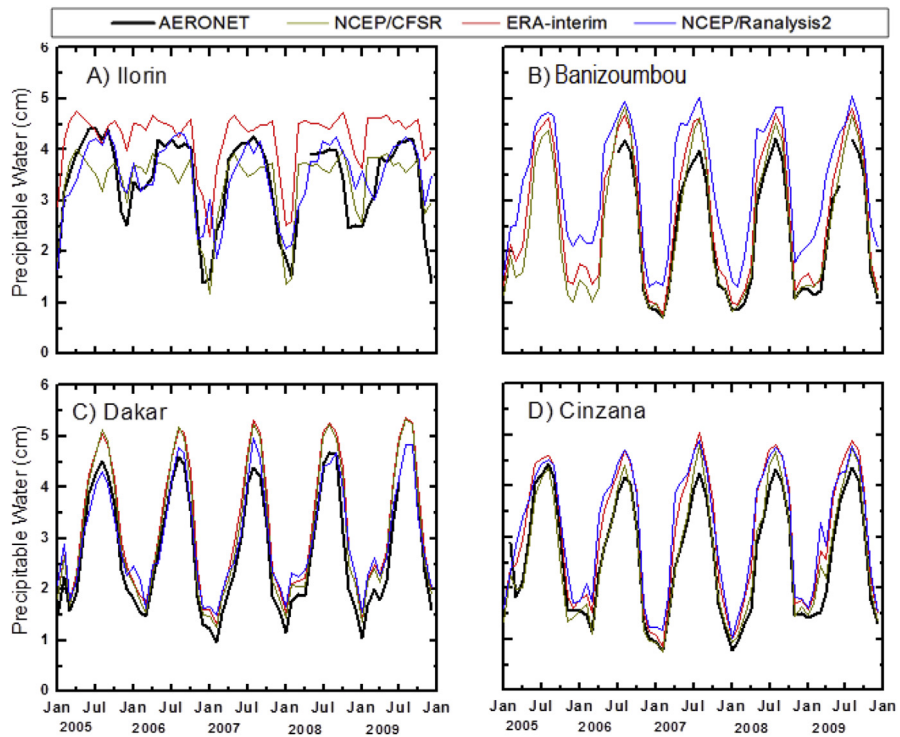

**Fig. 6.** Precipitable Water Vapor from ERA-i, NCEP CFSR, NCEP Reanalysis 2 and AERONET's CI-
MEL sun-photometer for 2005−2009 at (A) Ilorin, (B) Banizoumbou, (C) Dakar and (D) Cinzana.

**Table 1.** Statistical correlation between the PWV derived from the AERONET's
Sun-photometer and those derived from the NWP models.

| AERONET Station | NWP Models | $R^2$ | RMS error (cm) |
|---|---|---|---|
| Ilorin (8.3° N, 4.34° E) | NCEP Reanalysis II (7.5° N, 5.0° E) | 0.607 | 0.59 |
| | NCEP-CFSR (8.6° N, 3.8°E) | 0.638 | 0.42 |
| | ERA interim (8.25° N, 4.50°E) | 0.957 | 0.19 |
| Banizoumbou (13.54° N, 2.67° E) | NCEP Reanalysis II (12.5° N, 2.5° E) | 0.783 | 0.64 |
| | NCEP-CFSR (14.3° N, 1.9°E) | 0.982 | 0.19 |
| | ERA interim (13.50° N, 1.50°E) | 0.994 | 0.11 |
| Dakar (14.39° N, 16.96° W) | NCEP Reanalysis II (15.0° N, 17.5° W) | 0.861 | 0.43 |
| | NCEP-CFSR (14.3° N, 16.9° W) | 0.986 | 0.15 |
| | ERA interim (14.25° N, 17.25° W) | 0.974 | 0.20 |
| Cinzana (13.28° N, 5.93° W) | NCEP Reanalysis II (12.5° N, 5.0° W) | 0.856 | 0.50 |
| | NCEP-CFSR (12.4° N, 5.6° W) | 0.975 | 0.20 |
| | ERA interim (13.50° N, 6.00° W) | 0.985 | 0.17 |

 

**Table 2.** Statistical correlations between the derived empirical models, NWP models and the AERONET PWV. The grid points for the NWP models are NCEP Reanalysis 2 (for Ouagadougou: 12.5° N, 2.5° W; for Djougou: 10.0° N, 2.5° E), NCEP-CFSR (for Ouagadougou: 12.38° N, 1.88° W; for Djougou: 10.48° N, 1.88° E) and ERA interim (for Ouagadougou: 12.00° N, 1.50° W; for Djougou: 9.75° N, 1.50° E).

| Empirical Models | Validating Station | NWP Model | $R^2$ | RMS Error (cm) |
|---|---|---|---|---|
| $lnPWV = 0.055T_d + 0.068$ | Ouagadougou (12.20° N, 1.40° W) | AERONET | 0.858 | 0.48 |
| | | NCEP Reanalysis II | 0.809 | 0.56 |
| | | NCEP-CFSR | 0.812 | 0.58 |
| | | ERA interim | 0.937 | 0.33 |
| | Djougou (9.76° N, 1.60° E) | AERONET | 0.801 | 0.49 |
| | | NCEP Reanalysis II | 0.625 | 0.67 |
| | | NCEP-CFSR | 0.644 | 0.86 |
| | | ERA interim | 0.904 | 0.34 |
| $PWV = 39.29\left(\dfrac{e}{T}\right) + 0.25$ | Ouagadougou (12.20° N, 1.40° W) | AERONET | 0.857 | 0.48 |
| | | NCEP Reanalysis II | 0.808 | 0.56 |
| | | NCEP-CFSR | 0.818 | 0.57 |
| | | ERA interim | 0.935 | 0.33 |
| | Djougou (9.76° N, 1.60° E) | AERONET | 0.809 | 0.48 |
| | | NCEP Reanalysis II | 0.591 | 0.69 |
| | | NCEP-CFSR | 0.654 | 0.85 |
| | | ERA interim | 0.911 | 0.33 |
| $PWV = 13.44\left(\dfrac{RH}{T}\right) + 0.58$ | Ouagadougou (12.20° N, 1.40° W) | AERONET | 0.747 | 0.64 |
| | | NCEP Reanalysis II | 0.717 | 0.68 |
| | | NCEP-CFSR | 0.782 | 0.62 |
| | | ERA interim | 0.804 | 0.58 |
| | Djougou (9.76° N, 1.60° E) | AERONET | 0.797 | 0.49 |
| | | NCEP Reanalysis II | 0.648 | 0.65 |
| | | NCEP-CFSR | 0.775 | 0.68 |
| | | ERA interim | 0.837 | 0.44 |

Reanalysis 2 (about 25% higher than the CIMEL retrievals); this model overestimates precipitable water vapor during the dry season as well.

## 5. Conclusion

There is a large gradient in PWV over Africa during the months of January and July months with a strong reversal in the sub-Sahel from very dry conditions in winter to very humid ones in the summer. This seasonal variability explains the larger absolute differences between the observations and model estimates when the absolute values are high. The precipitable water vapor PWV, as estimated from ERA-Interim and NCEP/CFSR are found to be in a closer agreement with values retrieved from the AERONET's CIMEL sun-photometers in West Africa than the NCEP/Reanalysis 2 product (see Table 2) which was found to overestimate precipitable water vapor in all seasons by as much as 25 %.

AERONET precipitable water vapor data from the years 2004–2014 have been used to evaluate several empirical expressions based on conventional moisture parameters

 

more readily available in this region (i.e., relative humidity, ambient temperature, dew-point temperature, and vapor pressure). These empirical formulations have been found to perform reasonably well statistically; Eq. (4) has been found to be most appropriate for the estimation of precipitable water vapor in the sub-Sahel West Africa. Due to the critical importance of the sub-Sahel in climate research, it is of great interest to correctly estimate moisture parameters in climate models. As such, review of the presented empirical models, as more data is available, is suggested.

## Declarations

### Author contribution statement

### Funding statement

This research did not receive any specific grant from funding agencies in the public, commercial, or not-for-profit sectors.

### Competing interest statement

The authors declare no conflict of interest.

### Additional information

Data associated with this study has been deposited at www.aeronet.gsfc.nasa.gov/new_web/data.html.

### Acknowledgements

We wish to greatly thank the Principal Investigators (PIs) and the Site Managers for their efforts in establishing and maintaining AERONET stations at Ilorin, Cinzana, Banizoumbou, Dakar, Djougou and Ouagadougou. Much appreciation also goes to NOAA/OAR/ESRL PSD, Boulder, Colorado, USA (for making the NCEP Reanalysis data available through their website http://www.esrl.noaa.gov/psd/).